\documentclass[letter]{aa} \usepackage{txfonts,graphicx}
\usepackage{natbib}
\begin{document}
\title{A dust component $\sim$ 2 kpc above the plane in NGC 891}

\author{P. Kamphuis \inst{1} \and B. W. Holwerda \inst{2} \and R. J. Allen \inst{2} \and R. F. Peletier \inst{1} \and P. C. van
der Kruit \inst{1} } \institute{Kapteyn
Astronomical Institute, University of Groningen, Postbus 800, 9700 AV
Groningen, the Netherlands \and  Space Telescope Science Institute, 3700 San Martin Drive, Baltimore, MD 21218, USA}

 \abstract{The halo of NGC 891 has been the subject of studies for  more than a decade. One of its most striking features is the large asymmetry in H$\alpha$ emission.  In this letter, we will take a quantitative look at this asymmetry at different wavelengths for the first time.  }
{We  suggest  that NGC 891 is intrinsically almost symmetric and the large asymmetry in H$\alpha$ emission is mostly due to dust attenuation. We will quantify the additional optical depth needed to cause the observed asymmetry in this model.}{By comparing large strips on the North East side of the galaxy with strips covering the same area in the South West we can quantify and analyze the asymmetry in the different wavelengths.}{From the 24 $\mu$m emission we find that the intrinsic asymmetry in star formation in NGC 891 is small i.e., $\sim 30\%$.  The additional asymmetry in H$\alpha$ is modeled as additional symmetric dust  attenuation which extends up to  $\sim$ 40\arcsec (1.9 kpc) above the plane of the galaxy with a mid-plane value of $\tau$=0.8 and a scale height of 0.5 kpc}{}
 \maketitle

\section{Introduction}
\hspace*{0.5cm}NGC 891 is one of the best studied edge-on galaxies in the nearby universe. Its close proximity (9.5 Mpc, \citealt{1981A&A....95..116V}) and high inclination (89.8$\pm$0.5$\degr$, \citealt{2005MNRAS.358..481K}) make NGC 891 an excellent candidate for the study of the vertical structure of spiral galaxies. \\
\hspace*{0.5cm} The z-structure of the emission from NGC 891 has been studied at many wavelengths over the past $\sim$ 30 years. In bands such as the optical \citep{1981A&A....95..116V,1998A&A...331..894X} and the HI \citep{1979A&A....74...73S,1997ApJ...491..140S},  NGC 891 is almost completely symmetric. The  blue colors in the plane are  brighter on the the North East  side of the galaxy. This agrees with the idea of a trailing spiral structure with the dust behind the HII regions and the fact that the North East side is the approaching side  as first suggested by R. Sancisi (see, \cite{1981A&A....95..116V} and references therein).  The first sensitive H$\alpha$ studies \citep{1990A&A...232L..15D,1990ApJ...352L...1R} however,  observed a large  asymmetry between the North-East (NE) and the South-West  (SW) of NGC 891, extending up to large distances above the plane. H$\alpha$ is a tracer of ionized gas, often linked to star formation (SF from here on) because generally only young, massive stars produce enough ionizing photons to ionize large volumes of neutral gas.\\
\hspace*{0.5 cm}This asymmetry in H$\alpha$ could be an effect of dust attenuation affecting this tracer in the same way as the blue colors in the plane mentioned earlier. Even high above the plane ($\vert$z$\vert$= 2 kpc) dust structures can be located and identified \citep{2000AJ....119..644H} and it is natural to think  that attenuation by this dust on the SW side of NGC 891 can cause the asymmetry.\\ 
\hspace*{0.5cm}\cite{1978A&A....62..397A}  and \cite{1994A&A...290..384D} observed NGC 891 with high resolution and high sensitivity in radio continuum emission, finding the asymmetry observed in H$\alpha$  to be present to a lesser extent in  the distribution of radio continuum. \cite{1994A&A...290..384D}  suggest that this is the result of outflow of relativistic electrons produced as a by-product of SF in the disk. This and the fact that the asymmetry in H$\alpha$ extends well above the plane has led people to believe that it must be an intrinsic effect \citep{1994A&A...290..384D,2000AJ....119..644H,2007astro.ph..3363K}.\\
\hspace*{0.5cm} Recently a new SF tracer unaffected by dust attenuation has become available. \cite{2005ApJ...633..871C} show that 24 $\mu$m emission has a tight correlation with star formation.  Although this correlation does not work in the same way for all galaxies \citep{2006ApJ...648..987P}, the total Far Infrared Luminosity of NGC 891  (log(L(FIR))= 10.27 L$_{\odot}$) is so close to that of M51 (log(L(FIR))= 10.42 L$_{\odot}$, \citealt{2003AJ....126.1607S}) that this  correlation can be used for NGC 891. Based on this new SF tracer we will provide for the first time a quantitative multi-wavelength analysis of the asymmetry in the halo of NGC 891.

\section{Data}
\begin{figure*}[tbp]
\centering \includegraphics[width=14cm]{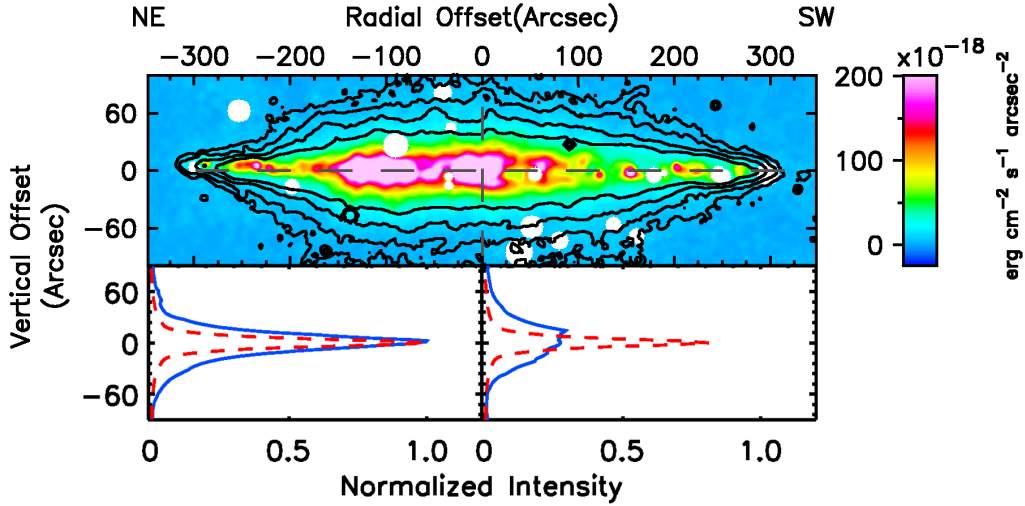}
\caption{Upper Panel: H$\alpha$ distribution of NGC 891 from \cite{
1990ApJ...352L...1R}  in units of 10$^{-18}$ erg cm$^{-2}$ s$^{-1}$ arcsec$^{-2}$ and convolved to the 6.4\arcsec resolution of the 24 $\mu$m.  Overlaid in black contours is the 24 $\mu$m emission at 3, 6, 12, 24 $\sigma$ levels
with $\sigma$=0.033 MJy sr$^{-1}$. Lower Panels: Intensity profiles parallel to the minor axis, averaged over 180\arcsec in the radial direction and 6.4\arcsec  in the vertical direction.  The profiles are normalized to the peak emission on the NE side. The panels show the NE (left panel) and SW (right panel) side of the galaxy. Solid blue line: H$\alpha$, Dashed red line:  24 $\mu$m.   Note that the difference between NE and SW in 24 $\mu$m is minimal. }\label{fig1} \centering
\end{figure*}

\hspace*{0.5cm}From the Spitzer Archive we retrieved the public mosaics (PBCD) of NGC 891 at 3.6 $\mu$m (IRAC) \citep{ Ashby08, Holwerda08} and at 24 $\mu$m (MIPS),  with the archive tool {\it Leopard}.\\
\hspace*{0.5cm} For the analysis of H$\alpha$ primarily the image of \cite{1990ApJ...352L...1R} is used.  An integrated map constructed from 
 a H$\alpha$ Fabry-P\'erot measurement \citep{2007astro.ph..3363K} is used to rule out any systematic effects which might be present in  the imaging. Since
the cube is calibrated on the image we will refer to the image of \cite{1990ApJ...352L...1R} in the discussion unless mentioned otherwise.\\
\hspace*{0.5cm}The Near and Far Ultra-Violet images of NGC 891 are retrieved from the GALEX public archive through the MAST website\footnote{http://archive.stsci.edu}. Unfortunately, the NUV is too crowded with foreground stars to
give any useful information on diffuse emission out of the plane of the galaxy.\\
\hspace*{0.5cm} The HI map and 21 cm continuum image are from the most recent observations of NGC 891 with the Westerbork Radio Synthesis Radio telescope  \citep{2007arXiv0705.4034O}. \\
\hspace*{0.5cm}Except for blanking stars in all the optical and IR images, no further reduction was performed on the images. The resolution of the optical and IR images was determined by fitting gaussian point spread functions to 6 stars in each image. The main characteristics of the images are summed up in Table \ref{tab1}.\\
\begin{table*}[tpb]

\centering
\begin{tabular}{llllll}

\hline
Name & (Central) Rest Wavelength & Type of emission & FWHM  & noise ($\sigma$) &\\
\hline
HI & 21 cm & Line & 30\arcsec & 0.261 &MJy sr$^{-1}$\\
21 cm & 21 cm & Continuum & 17\arcsec &  0.005 & MJy sr$^{-1}$\\
24 $\mu$m&  24 $\mu$m &Continuum&6.4\arcsec & 0.033 &MJy sr$^{-1}$ \\
3.6 $\mu$m& 3.6 $\mu$m& Continuum& 2.6\arcsec & 0.59 &MJy sr$^{-1}$\\
H$\alpha$ \citep{1990ApJ...352L...1R} &6563 $\AA$ & Line & 1.2\arcsec& 7.8$\times$10$^{-18}$ &erg cm$^{-2}$ s$^{-1}$ arcsec$^{-2}$\\
H$\alpha$ \citep{2007astro.ph..3363K} &6563 $\AA$ & Line & 2\arcsec&6.0$\times$10$^{-18}$ &erg cm$^{-2}$ s$^{-1}$ arcsec$^{-2}$\\
NUV &2275 $\AA$ &Continuum& 5.1\arcsec&0.003 &MJy sr$^{-1}$\\
FUV &1550 $\AA$ &Continuum& 4\arcsec &0.002 &MJy sr$^{-1}$\\
\hline
\end{tabular}
\caption{Summary of the characteristics of the different wavelength images.}\label{tab1}
 \end{table*}
\hspace*{0.5cm} Figure \ref{fig1} (upper panel) shows the distribution of  H$\alpha$
emission  overlaid with contours of the 24 $\mu$m emission. The lower panel of Fig. \ref{fig1} shows the intensity profiles parallel to the minor axis, averaged over 180\arcsec in the radial direction and  6\arcsec in the vertical direction. The profiles are normalized to the peak emission on the NE side.  The NE side of NGC 891 is shown in the left panel and SW side in the right panel. The solid line represent the H$\alpha$ and the dashed line the 24 $\mu$m. In Fig. \ref{fig1} the 24 $\mu$m profiles appear to be much narrower than the H$\alpha$ profiles. However, at low levels these profiles have large extended wings. These wings are still well above the noise level of the image but are hardly seen in the vertical profiles due to the large dynamical range of the emission. \\
\section{Results}
\begin{figure*}[tbp]
\centering \includegraphics[width=14cm]{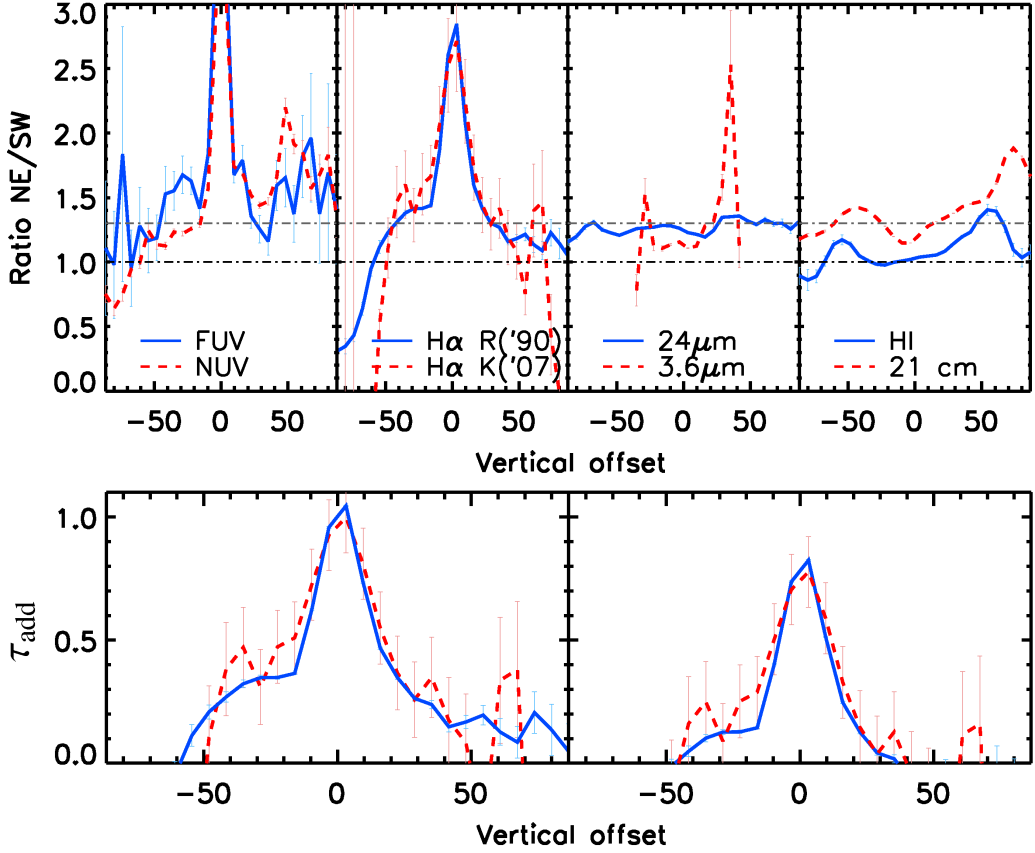}
\caption{Upper Panels: Ratio between the total emission (see text)
in the North East and the South West.  Left panel  FUV (Galex), NUV (Galex); Left middle panel H$ {\alpha}$ \citep{
1990ApJ...352L...1R, 2007astro.ph..3363K}; Right middle panel  24 $\mu$m
(Spitzer), 3.6 $\mu$m (Spitzer); Right panel HI, 21 cm continuum \citep{2007arXiv0705.4034O}.  All panels show two dot-dashed lines for (Gray) a ratio of 1 indicating complete symmetry and (Black) a ratio of 1.3 as found by \cite{1994A&A...290..384D} in the 6 cm radio continuum.
Lower Panels: Additional
$\tau$ between the spiral arms as derived from the
H$\alpha$. Solid blue lines as from \cite{1990ApJ...352L...1R}. Dashed red lines as from \cite{2007astro.ph..3363K}. Left Panel: assuming complete symmetry. Right Panel: assuming an intrinsic asymmetry as observed in the 24 $\mu$m.}\label{fig2} \centering 
\end{figure*}
\hspace*{0.5cm}We are  interested in determining what fraction of the SF in NGC 891 is intrinsically asymmetric, and whether the remainder  could be reasonably explained as extinction owing to dust at high $\vert$z$\vert$ in the galaxy.  We have averaged over large strips on either
side of the galaxy to obtain the  highest S/N possible. The strips
have a width of 180\arcsec son the major axis on either side and start at a
radial offset of 10\arcsec from the center so that effects from a central active
source will not contaminate our results. We are interested in
how the ratio between the emission in the NE and the SW changes with height in the different wavelengths. The resolution of the 24 $\mu$m (i.e., $\sim$ 6
\arcsec, see Table \ref{tab1}) is the lowest resolution in our different bands with the exception of the 21 cm data. Therefore we have set the height of the strips equal to one resolution element in the 24 $\mu$m image. The average intensity
of such a strip on the NE is then divided by the average
intensity of the strip on the SW with the same vertical
offset. This way we obtain a vertical profile for the ratio between NE
and SW with a resolution of 6\arcsec. These vertical profiles are
shown in the top half of Figure \ref{fig2}.\\
Figure  \ref{fig2} shows us that:    
\begin{itemize}
\renewcommand{\labelitemi}{$1. $}
\item The emission in 24 $\mu$m, 3.6 $\mu$m and HI is almost symmetric.
\renewcommand{\labelitemi}{$2. $}
\item H$\alpha$ and UV emission is asymmetric in the plane and declines to the same
asymmetry as 24 $\mu$m, 3.6 $\mu$m and HI above the plane. 
\end{itemize}
\hspace*{0.5cm} We see that the ratio of asymmetry is very different between the wavelengths which are affected by dust attenuation (e.g., H$\alpha$ and the UV) and the ones which are not (e.g., HI, 24 $\mu$m and 3.6 $\mu$m). Also, as we move away from
the plane of the galaxy the asymmetry becomes less pronounced in the bands affected by dust attenuation.\\
 \hspace*{0.5cm}The 24 $\mu$m and HI emission are not affected by dust
attenuation and the ratio between the NE and SW is constant with height in these bands. The ratios in the plane of the galaxy are 1.24, 1.03 and 1.19 for the 24 $\mu$m, HI and 21 cm continuum respectively. The ratio found in the 24  $\mu$m emission is very similar to the  ratio of 1.3 found in the radio continuum ($\lambda=6$ cm) by \cite{1994A&A...290..384D}. Therefore, we can consider this ratio as the intrinsic asymmetry of SF in NGC 891. We model the rest as the effects of dust attenuation. \\
\section{Discussion}
\begin{figure}[tbp]
\centering \includegraphics[width=8cm ]{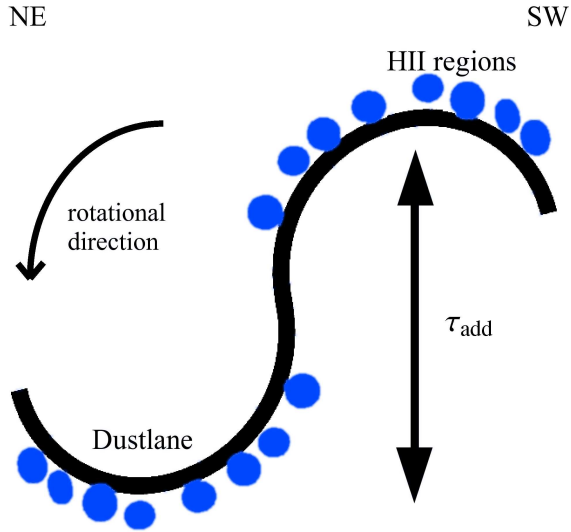}
\caption{Face-on view of our model for a symmetric NGC 891. The black S shape symbolizes the dust lane in the spiral arms and the blue areas the HII regions. The straight double headed arrow shows the difference in line of sight between the North East side and the South West side. The curved single headed arrow indicates the direction of rotation in NGC 891.}\label{fig3} \centering 
\end{figure}

\hspace*{0.5cm}Given the symmetry of the gas (HI),  non-attenuated stellar (3.6 $\mu$m) and SF (24 $\mu$m) emission (Fig \ref{fig2}, right middle panel), all tracing different components of NGC 891, we assume that NGC 891 is intrinsically symmetric in the plane. Furthermore, we assume that the H$\alpha$ above the plane is related to the SF in the plane. This means that the H$\alpha$ above the plane is also symmetric and situated above the HII regions in the plane. We constructed a model  based on the trailing spiral arm idea, which means that the HII regions are located in front of the dust lanes of the spiral arms. Then, because the SW is the receding side of the galaxy, on the NE we would be
looking first at the HII regions and then the dust in the arm but on the
SW side our line of sight to the HII regions passes through the arm first. Figure \ref{fig3} shows the face-on view of our model of NGC 891. \\ 
\hspace*{0.5cm}For the attenuating dust in the halo we assume an isotropic distribution (e.g., not specifically localized above the SF regions). It is already shown that the distribution of attenuating  dust and  ionized gas are unlikely to trace the same component \citep{2000AJ....119..644H}.  Furthermore, on the SW side tendrils, chimneys and shells are identified by \cite{2000AJ....119..644H}. It is unlikely that these structures would be  backlit by the stellar disks if they resided in the back of the galaxy.\\
\hspace*{0.5cm} Therefore, in our model, the asymmetry in the H$\alpha$ could easily be explained by dust in the halo. If the spiral arm on the SW side would be further away from us than the spiral arm on the NE, then the sight line through the dust toward the H$\alpha$ on the South West of the galaxy would be much longer. From this  idea we can easily calculate the additional attenuation between the two arms. The lower panel in Figure \ref{fig2} shows  the additional optical depth in the halo between the two spiral arms. $\tau_{\rm add}$ is calculated by assuming that the H$\alpha$ emission from the NE side of the galaxy is equal to the intrinsic emission on SW side and that all the differences are caused by dust attenuation. This is shown in the lower left panel of Figure \ref{fig2} where the solid black line is  from the image \citep{1990ApJ...352L...1R} and the solid gray line is from the integrated moment map \citep{2007astro.ph..3363K}.  The dust extends easily to 50\arcsec (2.3 kpc) above the plane.  Even if we take the small asymmetry of the 24 $\mu$m in the plane into account (Fig \ref{fig2}, lower right panel)  additional attenuation is present up to $\sim 40$\arcsec (1.9 kpc). \\
\hspace*{0.5 cm} Fitting a single exponential to the inner part ($\vert$z$\vert <$ 20\arcsec) of these optical depth profiles, we obtain a scale height for the dust of 0.67$\pm$0.10 kpc if the galaxy is perfectly symmetric and of 0.46 $\pm$ 0.10 kpc if the intrinsic asymmetry is $\sim$ 1.3.   This is similar to the scale height of the old stellar population  ($z_s$=0.56 kpc in 3.6 $\mu$m, \citealt{Holwerda08}) in NGC 891. This is consistent with the results for low mass galaxies, where the vertical dust distribution has a scale height somewhat larger than that of the main sequence stars \citep{Seth05}. However, it does complicate the picture for high mass galaxies, where previously it was thought that the ISM completely collapsed into the typical dustlanes \citep {2004ApJ...608..189D}.  \\
\hspace*{0.5 cm}The scale heights  were calculated from profiles with a vertical resolution of 1\arcsec to ensure that we had enough points to fit. These scale heights can be considered a lower limit to the real scale height of the dust since the scale height calculated in this way is a combination of the dust scale height and the H$\alpha$ distribution.  \\
\hspace*{0.5 cm}Higher up ($\vert$z$\vert >$ 20\arcsec) the profile hints at a second component on the East side (e.g., negative vertical offset) with an even larger scale height .
 This would imply that there is a dust halo with a scale height $\geq$ 0.6 kpc.  Thus, the extraplanar dust has a much larger scale height than previously assumed  \citep{Xilouris99, 2004ApJ...608..189D, Seth05,Bianchi07}.  The quality of the data, in our opinion, is not good enough to fit a scale height to this second component. \\ 
\hspace*{0.5 cm} Moreover, if we fit the 24 $\mu$m emission with a double exponential we find the "normal" small scale height for the region close to the plane ($z_ {24  \mu{\rm m}}$=0.22 $\pm$ 0.05 kpc) but a much larger second component ($z_ {24  \mu{\rm m}}$=1.3 $\pm$ 0.1 kpc) above 30\arcsec (1.4 kpc). These values are no more than indicative since we did not correct for the wings of the PSF in 24 $\mu$m image.\\
\hspace*{0.5 cm} The fact that the inferred values for $\tau_{\rm add}$ are very reasonable, implies that the ionized gas in the halo is closely related to the star formation in the plane of the galaxy. Moreover, if our model is the correct view of NGC 891, the ionized gas and the attenuating dust follow very different distributions in the halo: the dust is equally distributed around the disk while the ionized gas remains above the position it was initially ejected for a long time.  

\section{Conclusions}
\hspace*{0.5 cm} For the first time we quantitatively analyze the asymmetry between the North East and the South West side of the NGC 891 at several wavelengths, thus tracing different components from the galaxy. We determine the ratio of asymmetry by dividing large strips on the North East side by the same area on the South West side of the galaxy (Fig. \ref{fig2}, upper panel).  From this analysis it is immediately obvious  that star formation tracers affected by dust attenuation (e.g., H$\alpha$, UV) show a much larger asymmetry than star formation tracers unaffected by dust  attenuation (24 $\mu$m, radio continuum). We also see that the other components in the galaxy, like the old stellar population (3.6  $\mu$m) and the neutral gas (HI) have only a small asymmetry. \\
\hspace*{0.5 cm}Assuming that 24 $\mu$m emission from star forming regions dominates in the plane, the 24 $\mu$m emission has a direct correlation with star formation \citep{2005ApJ...633..871C}.  The small asymmetry we find in the 24 $\mu$m emission is confirmed by the result found in the radio continuum \citep{1994A&A...290..384D}, which leads us to believe that 24 $\mu$m is indeed a good tracer of star formation in NGC 891. Therefore, the asymmetry in H$\alpha$ is most likely caused by dust in and above the plane. \\
\hspace*{0.5 cm} From this information we construct a simple symmetric model for NGC 891. This model is mainly based on the trailing spiral arm idea where the HII regions are located in front of the dust lanes (Fig \ref{fig3}). We assume that the emission from the ionized gas is located in and above these HII regions but that the dust distribution in the halo is much more isotropic over the disk. This would lead to a longer sight line through the dust towards the H$\alpha$ emission on the South West side of the galaxy.\\ 
\hspace*{0.5 cm} From this model we can derive the additional attenuation ($\tau_{\rm add}$) in the halo between the spiral arms (Fig \ref{fig2}, lower panel). We see that at heights $\sim$ 40\arcsec  (1.9 kpc) the additional attenuation of H$\alpha$ becomes negligible. This is either caused by the absence of dust at this height or by the distribution of the ionized gas becoming such that our  lines-of-sight toward the ionized gas become equal on both sides. This last scenario, implies that the attenuation in the halo is even higher above the plane than derived from our simple model.\\
\hspace*{0.5 cm} The opacity values calculated in this letter only represent the additional opacity between the spiral arms (see fig. \ref{fig3}, straight arrow. Thus, these values can be considered as a lower bound on the true opacity through the whole galaxy as a function of vertical height. \\
\hspace*{0.5 cm}Common wisdom still has it that extinction in the disk is predominantly in the plane itself. Typical scale heights are much smaller than the stellar ones \citep{Xilouris99,Bianchi07} or similar \citep{Seth05}. This paper shows that there is a second spatial component to the dusty ISM with a much more extended scale height.\\
\hspace*{0.5 cm}If our view of NGC 891 is correct then the unattenuated H$\alpha$ emission above the plane is closely correlated with star forming regions in the plane. This implies that the ionized halo gas is mostly brought up from the plane through processes related to star formation. Furthermore the distribution of the attenuating dust above the plane is much more isotropic, and therefore more stable, than that of the ionized gas. This is probably caused by the ionized gas cooling down and recombining  before it reaches this stable configuration.

\begin{acknowledgements}
We would like to thank R. Rand for kindly providing the H$\alpha$ image of NGC 891. T. Oosterloo for the HI map and the 21 cm continuum image. The J.C. Kapteyn fund and STScI  financed P. Kamphuis' work visit to Baltimore.
\end{acknowledgements}

\bibliographystyle{bibtex/aa}   \bibliography{../ref}
\end{document}